\documentclass[12pt, letterpaper]{article}
\usepackage[utf8]{inputenc}
\usepackage[super]{natbib}
\usepackage{amsmath,amssymb,amsfonts}

\usepackage{algorithm}
\usepackage{algorithmic}
\usepackage{graphicx}
\usepackage{textcomp}
\usepackage{xcolor}
\usepackage{eucal}
\usepackage[T1]{fontenc}
\DeclareMathOperator*{\argmax}{arg\,max}
\DeclareMathOperator*{\argmin}{arg\,min}
\usepackage{authblk}
\usepackage[font=bf]{caption}
\usepackage[hyphens]{url}
\usepackage{hyperref}
\hypersetup{breaklinks=true}

\usepackage{lastpage}
\usepackage{fancyhdr}
\title{\fontsize{14}{15}\bfseries 
DRLDO: A novel DRL based De-Obfuscation\\
System for Defense against Metamorphic Malware}

\author{
Mohit Sewak\\
\texttt{Security \& Compliance Research, Microsoft, India}\\
\texttt{mohit.sewak@microsoft.com}
\and
Sanjay K. Sahay, Hemant Rathore\\
\texttt{Department of CS\&IS, BITS Pilani, Goa, India}\\
\texttt{\{ssahay, hemantr\}@goa.bits-pilani.ac.in}
}

\date{} 

\begin{document}
\maketitle

\begin{abstract}
\boldmath
In this paper, we propose a novel mechanism to normalize \textit{metamorphic} and \textit{obfuscated} malware down at the \textit{opcode level} and hence create an advanced metamorphic malware de-obfuscation and defense system. We name this system as DRLDO, for \textbf{D}eep \textbf{R}einforcement \textbf{L}earning based \textbf{D}e-\textbf{O}bfuscator. With the inclusion of the DRLDO as a sub-component, an existing Intrusion Detection System could be augmented with defensive capabilities against `zero-day’ attack from obfuscated and metamorphic variants of existing malware. This gains importance, not only because there exists no system till date that use advance DRL to intelligently and automatically normalize obfuscation down even to the opcode level, but also because the DRLDO system does not mandate any changes to the existing IDS. The DRLDO system does not even mandate the IDS' classifier to be retrained with any new dataset containing obfuscated samples. Hence DRLDO could be easily retrofitted into any existing IDS deployment.
We designed, developed and conducted experiments on the system to evaluate the same against multiple-simultaneous attacks from obfuscations generated from malware samples from a standardized dataset that contain multiple generations of malware. Experimental results prove that DRLDO was able to successfully make the otherwise un-detectable obfuscated variants of the malware detectable by an existing pre-trained malware classifier. The detection probability was raised well above the cut-off mark to $0.6$ for the classifier to detect the obfuscated malware unambiguously. 
Further, the de-obfuscated variants generated by DRLDO achieved a very high correlation (of $\approx 0.99$) with the base malware. This observation validates that the DRLDO system is actually learning to de-obfuscate and not exploiting a trivial trick.
\end{abstract}

\section{Introduction}\label{introduction}
Metamorphism provides malware an effective mechanism of evading an Intrusion Detection Systems (IDS). The different metamorphic variants of a metamorphic malware are functionally equivalent, but their internal structures or source codes may differ. Code obfuscation methods like dead-code insertion are widely used for developing metamorphic malware \cite{metamorphic_detection}. The problem of de-obfuscation (/normalization) at the opcode level, when transformed into a Markov Decision Process (MDP), leads to a Reinforcement Learning (RL) task that involves high cardinality action space. Most of the popular RL or Deep Reinforcement Learning (DRL) \cite{Sewak-DRL} agent algorithms like the Deep Q Networks (DQN) \cite{DQN_Atari}, or even its enhanced variants like the Double \cite{Double_DQN} or Dueling Deep Q Networks \cite{Dueling_DQN} could not efficiently learn an optimal policy under such constraints \cite{DDPG}. These systems also suffer from sample-inefficiency when applied to complex MDP. Such MDP(s) requires sophisticated DRL agents to effectively learn the policy and find effective and sample-efficient solutions to the given MDP. Training such DRL agents also requires sufficient data to ensure stable and robust learning. 
In this paper, we present a novel method for de-obfuscation of advanced metamorphic, oligomorphic, and polymorphic malware using Deep Reinforcement Learning (DRL). We designed and developed a working system named \textbf{DRLDO} (\textbf{D}eep \textbf{R}einforcement \textbf{L}earning based \textbf{D}e-\textbf{O}bfuscator). The objective of the DRLDO system is to train DRL agents that can perform the task of transforming/ normalizing/ de-obfuscating the obfuscated malware's opcode sequence features. Doing so a DRLDO based system could be fitted inside an IDS before the malware-classification system. The DRLDO system would de-obfuscated the incoming candidate file's features sufficiently so as the IDS' classification-system could detect these otherwise un-detectable feature-vectors correctly (as malicious) upon de-obfuscation without mandating any re-training or re-calibration of the IDS. Additionally, it is desirable that the transformed feature-vectors as generated by the DRLDO are similar (demonstrate high correlation between feature-vectors) to that of the original malware's to ensure that the DRLDO system is also compatible with IDS incorporating multinomial-classification-system and other advanced sub-systems that requires to identify the specific strain of malware for further processing. 
We conducted experiments with the developed system to gauge its effectiveness against multiple-simultaneous attacks from different obfuscated variants of malware extracted from a standardized malware dataset \cite{Malicia}. Experimental results prove that the DRLDO system could normalize/ transform/ de-obfuscated the obfuscated malware's feature-vectors such that these malware could subsequently be correctly detected by existing IDS (which had failed to detect the metamorphic instance correctly) without mandating any re-training.
Another significant achievement of the DRLDO system is that the de-obfuscations produced from it resulted in an opcode frequency vector which were very similar to that of the original malware variant's opcode frequency vector. 
This observation adds further credibility to the assertion that the DRLDO system is actually trying to de-obfuscate the malware's opcode frequency vector (of any junk opcode/ instruction insertion) and not just finding a trivial trick to ensure that IDS start detecting the metamorphic instances as malicious. 
The remaining of the paper is organized as follows. In section \ref{related-work} we discuss some of the related work. To the best of our understanding DRLDO is the first DRL based system for creating opcode level de-obfuscation, so no related work could be found in this specific area. But we present related work in other areas of generative malicious network traffic creation using RL and generative (non-RL) machine learning for malware creation. 
Next, in section \ref{process-flow}, \ref{rl-environment} and \ref{rl-agent} we cover the details of the design of our system, the custom reinforcement learning environment that we created for this setup, and the DRL that we used. Next, in section \ref{results} we discuss the different experiments conducted with the DRLDO system and the corresponding results. Finally, we discuss the results in section \ref{discussion} and conclude the paper in section \ref{conclusion}.

\section{Related Work} \label{related-work}
Recently there has been a lot of interest to improve the performance of IDS against unseen intrusions especially in network traffic and botnet attack areas. Most of the initial efforts in this regard were inspired by the Generative Adversarial Networks (GAN) as formed by the combination of two different Convolution Neural Networks (CNNs) \cite{Sewak-CNN} where the detection feedback from one network called the `Discriminator CNN' (denoted as $'\mathcal{D}'$) is used to train the other CNN network called the `Generator CNN' (denoted as $'\mathcal{G}'$). 
With the initial inspiration being drawn from the GAN networks of underlying CNN architecture, some aspects of network traffic were converted into a similar CNN map on which a GAN style methodology could be adopted to produce the necessary `generative' perturbation by the `Generator CNN' to create samples that the `Discriminator' CNN could not identify correctly. 
This problem could be formulated as a mini-max contest between $'\mathcal{D}'$ and $'\mathcal{G}'$, where, $'\mathcal{D}'$ is trying to maximize the cross-entropy error of detection of samples produced by $'\mathcal{D}'$ using $'\mathcal{G}'$ and $'\mathcal{G}'$ is trying to minimize it attractively. This could be defined mathematically as the optimization function in equation \ref{eq:GAN}:
\begin{equation}
    \min_\mathcal{G} \max_\mathcal{D}   \mathbb{E}_p(x)\log{\mathcal{D}(x)}  +   \mathbb{E}_p(z)\log{(1-\mathcal{D}(\mathcal{G}(z)))}
    \label{eq:GAN}
\end{equation}
Since this is a differentiable equation, it could be optimized to train and weight of the $'\mathcal{D}'$ and $'\mathcal{G}'$ networks that converges gradually and then the samples from $'\mathcal{D}'$ are used for the intended purpose. 
To create an undetectable malicious entity (file or network traffic), the network $'\mathcal{D}'$ could be replaced by the corresponding IDS's classifier's approximation function. Such approximation function could be any Deep Learning (DL) network. Most of these systems work to create generative data distributions that supposedly mimic an undetectable malicious entity (file or network traffic). While using such a system to create malicious traffic/ botnet, the $'\mathcal{D}'$ network could be replaced by the corresponding IDS's classifier's associated trained DL network, to learn to create perturbations in the distribution of existing (malicious) data and creating new (malicious) data distributions that is unknown to the existing IDS ($'\mathcal{D}'$ network). 
Some prominent example of such approach could be MalGAN \cite{MalGAN}, IDSGAN \cite{IDSGAN}, and by Usama et. al. \cite{UsamaGAN}. 
There are some non-GAN approaches which are based on differentiable objective functions, these are typically known as 'Gradient attack' based approaches. An example of a similar implementation is the Fast Gradient Sign Method (FGSM) \cite{GrosseFGSM}.
Additionally, there are some other non-differentiable objective based methods in the area of Reinforcement Learning \cite{Anderson_PE_RL}, and \cite{Wu_BotNet_RL} that have been explored.
These methods are designed with an underlying assumption that the distribution of the new data generated from the $'\mathcal{G}'$ network of these systems is significantly different form the ones that the IDS ($'\mathcal{D}'$ network) is trained on. Thus, on retraining the IDS on the data coming from this new distribution (in combination with their original training data) can improve the overall IDS system's response against an actual new attack for which the real data does not exist. But as pointed out in \cite{MalGAN} these methods, especially the ones based on GANs make re-training of IDS ineffective, and others may even make the IDS over-fit \cite{Anderson_PE_RL} if trained on such generated data thus reducing their effectiveness.
Therefore, it is not sufficient to just create a system that could generate malware samples that the IDS could not detect and hope that using these samples the IDS could be improved substantially in its ability to avert attacks from unknown intrusion attacks especially those by the complex obfuscations of existing malware. Also, it is not optimal to retrain the IDS with hypothetical data distributions generated with such perturbations, as this increases the risk of decreasing the effectiveness of the IDS on the actual existing malware detection. 
The above observation necessitates that instead of modifying the training of the existing IDS which is working well on the original/ un-obfuscated variants of the malware, the obfuscated variants of the threats need to be normalized to bring them close to the actual variant that the IDS was earlier trained upon and hence could probably detect.

\section{DRLDO Process Flow}\label{process-flow}
\begin{figure*}[!htbp]
    \centering
    \includegraphics[width=5in]{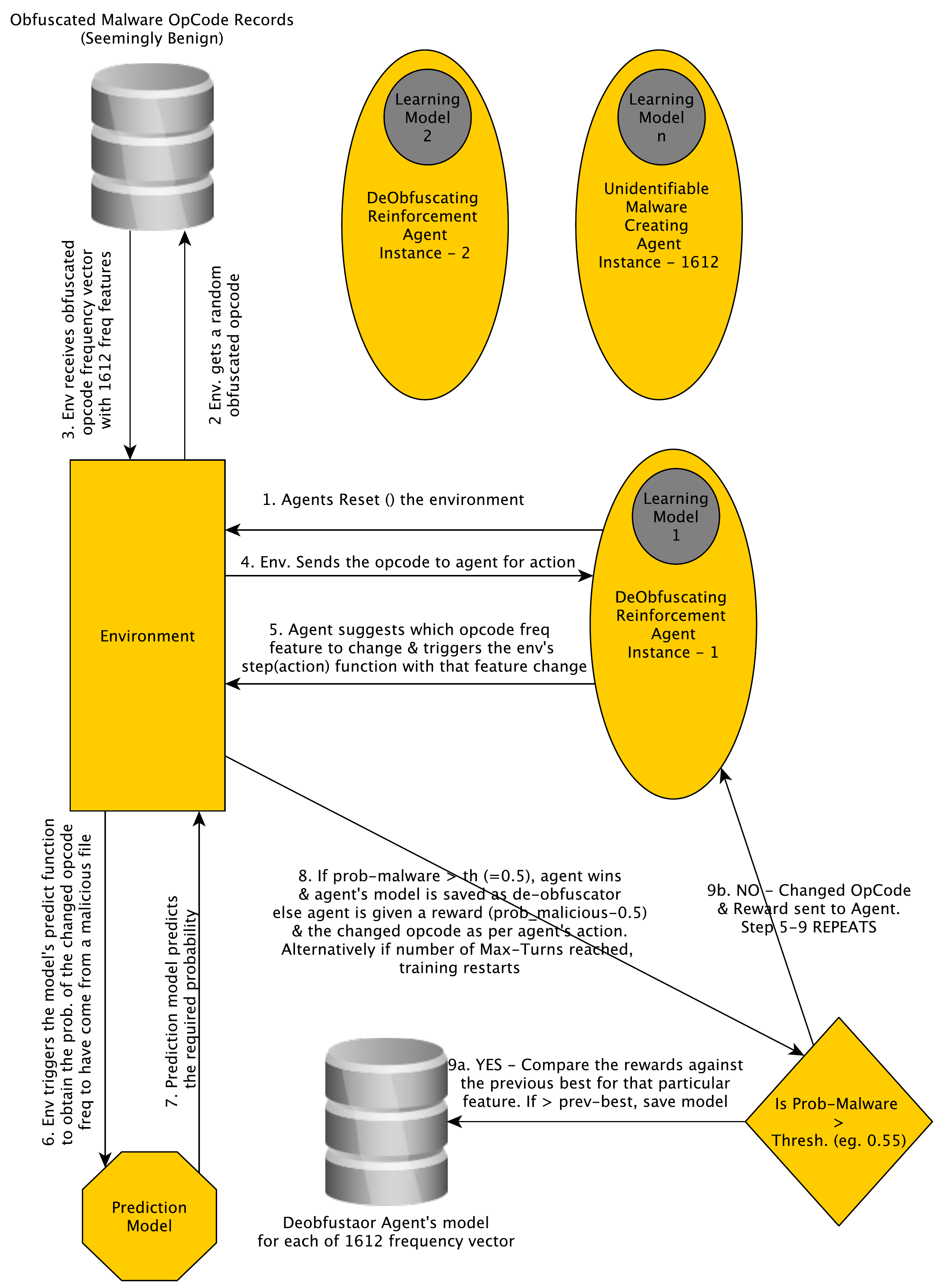}
    \caption{DRL based De-Obfuscation System - Process Flow.}
    \label{fig:process_flow_adversarial}
\end{figure*}

The process flow for the design of the DRLDO system is shown in figure \ref{fig:process_flow_adversarial}. It has broadly 4 sub-systems, namely: 
\begin{itemize}
    \item the obfuscated opcode repository and associated obfuscation generation system like the ADRLMMG (the detailed coverage of this system is out of scope of this paper),
    \item a pre-trained opcode frequency feature-vector generator and classification system (sub-components of existing IDS),
    \item a custom malware de-obfuscation training reinforcement learning environment,
    \item and a DRL agent compatible with both the custom environment and also the type of reinforcement learning problem at hand (conceptually and mathematically).
\end{itemize}
In the figure \ref{fig:process_flow_adversarial} shown, the DRL agent interacts with the environment to train against \textit{episodic} tasks. These tasks comprise of altering the presented opcode frequency vector. The opcode frequency vector is altered to achieve a net reduction of some of the selected opcode instructions, thus mimicking actions opposite to that in popular obfuscation techniques (i.e. of adding junk instructions). In each training step of every training episode, the environment receives and instruction from the DRL agent, and then transforms the opcode frequency structure as per the received instruction. The environment then uses the attached IDS to obtain the probability of the thus transformed opcode frequency vector to have come from a malicious file ($\mathbb{P}_{malicious}$). A decent increment in $\mathbb{P}_{malicious}$ from the initial state malicious detection probability (i.e. $\mathbb{P}_{malicious} < $ 0.5) of opcode frequency to a level where the classifier could subsequently detect the file instance as malicious (i.e. $\mathbb{P}_{malicious} > $ 0.5).
The DRL agents train over multiple such training episodes to update and refine an action policy that could be applied on any obfuscated malware strain to de-obfuscate it. The trained DRL agent is used to create and store opcode frequency vectors representing abstractions that could be identified by existing detection systems as malware even when they could not identify the obfuscated variants of these malware in a situation typically posed under a zero-day attack. So, this system essentially works to negate a zero-day attack otherwise possible by using undetectable obfuscated instances of different malware variants.
Multiple such DRL agents could be created and trained with varying degree of dissimilarity from other DRL agents trained on the same environment and subsequently producing probable de-obfuscation variants of the same malware that are dissimilar from each other at different levels. Such dissimilarities could range from the extremes of changing the underlying algorithm of the complete agent to just changing the random number seed of various instances of the same DRL agent.
Working at the opcode level the process design is universal and could be implemented on malware of all popular systems ranging from Android, Windows, clients to Linux and Windows based servers. In production, this system could be implemented on any incoming file to these systems without having a prior assumption of the benign/ malicious nature of the file or its obfuscation status. Coupled with both the original opcode frequency vector and the probable de-obfuscated opcode frequency, either the existing classification system's detection potential could be enhanced, or more powerful single stage or hierarchical detection system be made.

\subsection{Training Data and IDS used}\label{sec:training-data}
The obfuscated (metamorphic) opcode frequency feature-vector of existing malware were produced by the \textbf{ADRLMMG} system which is an \textbf{A}dversarial \textbf{D}eep \textbf{R}einforcement \textbf{L}earning based \textbf{M}etamorphic \textbf{M}alware \textbf{G}enerator. The ADRLMMG and the DRLDO systems are isolated systems and they do not share weights and are unaware of the other's existence. The ADRLMMG system creates obfuscations of existing identified malware and stores the resultant opcode frequency vector of these obfuscations as the training data for any downstream de-obfuscation system like the DRLDO.
The ADRLMMG system also maintains a dictionary of the original malware's identity and its opcode frequency feature-vector in a separate metadata for performance assessment of the downstream de-obfuscation system. The ADRLMMG system aims to mimics obfuscation techniques like junk instructions/ code insertion and also to preserves the functionality of the original malware and hence it uses strictly opcode additive action policy. 
Obfuscations of multiple malware and benign files were produced. The obfuscations of the malware that could not be identified by the existing IDS satisfactorily were retained and the remaining discarded. The original opcode frequency vector of each file was preserved for comparison with that of the final obfuscated variant as produced by the ADRLMMG and the resultant de-obfuscated version of each of the obfuscated variant as produced by the DRLDO systems.
We choose the IDS system (including the pre-processing, feature selection and transformation and the classification sub-systems) as developed by \cite{sewak-ARES, SewakSNPD, rathore2018malware} to augment it with Zero-Day-Defense \cite{sahay2020evolution} capabilities against metamorphic malware attack using the DRLDO system. The selected IDS had claimed to provide the best performance (with a combination of the highest accuracy accompanied with a very low false positive rates) over a standardized malware data \cite{Malicia} with mixed types and generation of malware. 
For the obfuscated malware samples, we used the same data on which the given IDS was trained (and claimed the associated performance). Since the chosen IDS is trained on the malware in this dataset and also claims to have performed well detecting these malware, we can safely assume all the malware in this dataset as `existing' malware that are `known' to the chosen IDS. 
We use the ADRLMMG system to produce multiple obfuscated variants of this data. Multiple obfuscated variants of each `existing' malware could be generated. Each obfuscated variant thus produced was screened against the chosen IDS to ensure that it is incorrectly identifying the obfuscations of malware as non-malicious ($\mathbb{P}_{non-malicious} > 0.5$) before using these obfuscated samples in our experiments with the DRLDO system and obtaining the corresponding results.

\subsection{Preserving existing IDS and File Functionality}
\begin{figure*}[htb]
    \centering
    \includegraphics[width=1\linewidth, height=2.5in]{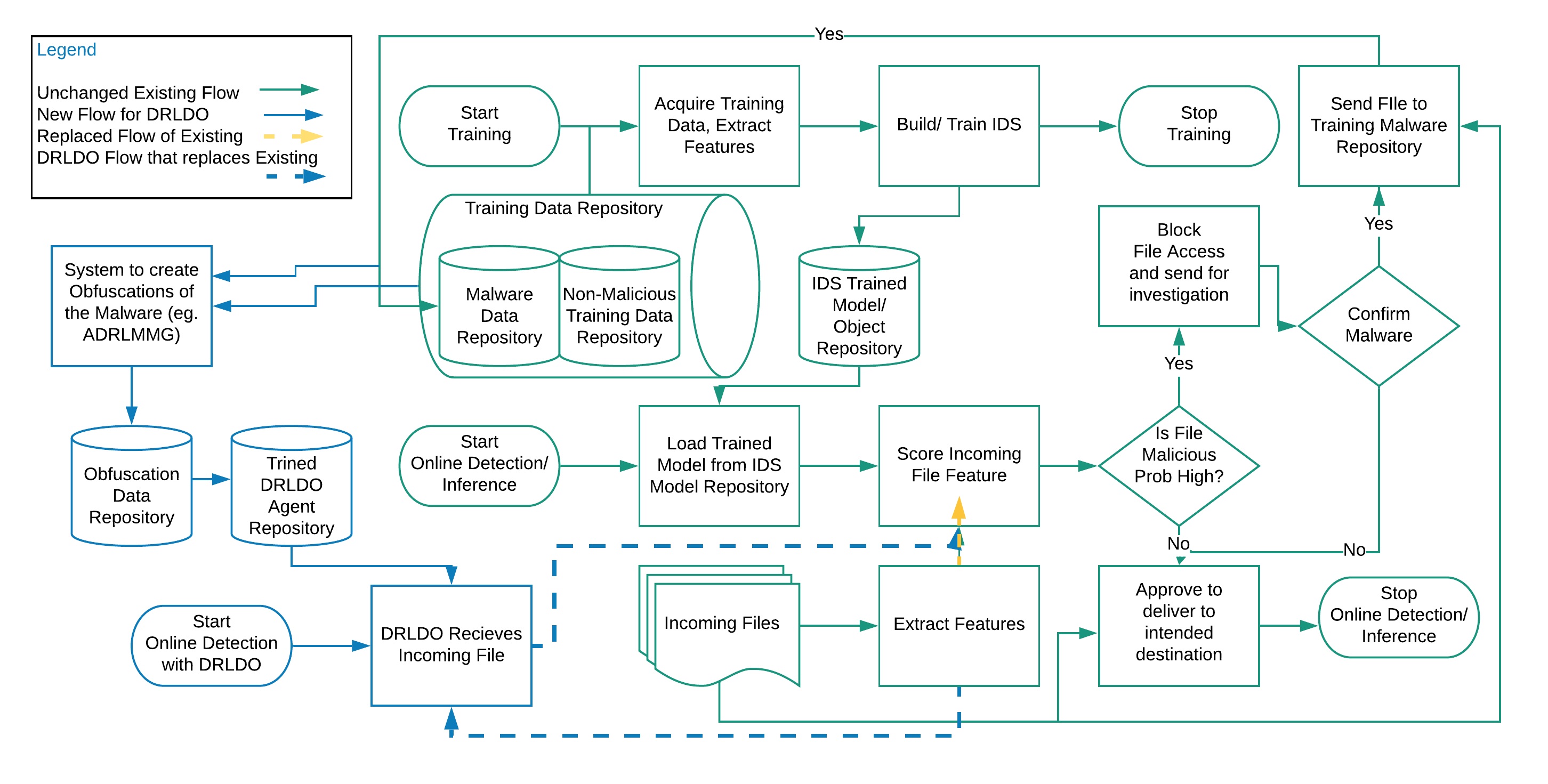}
    \caption{Preserving Functionality: DRLDO mandates no changes in the IDS not even retraining.}
    \label{fig:process_flow_preserving_functionality}
\end{figure*}

The figure \ref{fig:process_flow_preserving_functionality} shows the existing workflow of any IDS system that could cater to both offline and online detection requirements either in batch or in real-time mode. Additionally, there are additional components marked that are required for the enablement of the DRLDO system to augment existing IDS with metamorphic malware detection capabilities.
As could be found in this process, the existing components like the existing IDS setup, its related training workflows,  the associated training data etc. does not require any change. The IDS do not even need to be retrained to accommodate any obfuscation data. 
The only change that is required is just having file's feature-vector (as extracted by the IDS's feature generator) transformed by the DRLDO system before sending it to the IDS's classifier for detection. Besides this single change in the deployment architecture, the entire deployment setup remains the same, thus preserving the functionality and also the training, scoring and the deployment setups of the existing IDS. Similarly, the associated treatment and the functionality of the files passing through the system is also preserved.

\section{Custom Reinforcement Learning Environment}\label{rl-environment}
The environment serves a major role in reinforcement learning. As illustrated in algorithm \ref{algo:rl_env} its role is to present a current state to the agent to act upon, and then subsequently give it an appropriate  reward and the corresponding next state to the agent. The current\_state, action, reward, next\_state cycle continues until a terminal state is reached (for an episodic task) or until a predestined number of steps are completed. On reaching such scenario, the environment resets, itself and re-instantiates any default state and other necessary variables.

\begin{center}
\begin{algorithm}[b!]
\caption{Custom RL Environment Algorithm.}
    \begin{algorithmic}
        \REQUIRE RESET instructions from agent
        \ENSURE IDS, MalwareDateRepository is attached
        \STATE $ index \leftarrow random [1, \mathcal{N}_{MalwareDataRepository}]$
        \STATE STATE $\leftarrow IDSFeatureGenerator(index)$
        \STATE REWARD $\leftarrow 0$
        \STATE IsCOMPLETE $\leftarrow$ False
        \STATE $return \leftarrow (\text{STATE}, \text{REWARD}, \text{IsCOMPLETE})$
        \WHILE{$Turn \le MaxTurns \lor \neg IsCOMPLETE$}
            \STATE $RESPONSE \leftarrow \text{Agent(ACTION)}$
            \STATE $\text{NEW STATE} \leftarrow STATE(ACTION)$
            \STATE $\mathbb{P}_{malicious} \leftarrow IDSDetector(NEW STATE)$
            \IF{$\mathbb{P}_{malicious} \geq \mathbb{P}_{MalThreshold}$}
                \STATE $\text{REWARD} \leftarrow \text{REWARD}+Reward_{Probability}+Reward_{Victory}$
                \STATE $IsCOMPLETE \leftarrow True$
            \ELSE
                \STATE $\text{REWARD} \leftarrow \text{REWARD}+Reward_{Probability}-Penalty_{Turn}$
                \STATE $N \leftarrow n$
            \ENDIF
            \STATE $return \leftarrow (\text{STATE}, \text{REWARD}, \text{IsCOMPLETE})$
        \ENDWHILE
    \end{algorithmic}
    \label{algo:rl_env}
\end{algorithm}
\end{center}

\subsection{The structure of `State'}\label{env-state}
The state in our experiment is comprised of a vector of whole numbers ($ S \in \mathbb{W}^{|{opcode}|}$) corresponding to each of the unique opcode frequency for a given opcode in a file. We use the same unique opcode set as used by Sewak et. al. \cite{SewakSNPD}. We also use the same IDS which had produced the best performance and as used in their work. Their system claimed an accuracy of 99.21\%with a False Positive Rate of 0.19\% on the Malicia dataset \cite{Malicia} which is to our information the best performance achieved on this standardized malware dataset till date. The opcodes generated are also from the Malicia dataset. We obfuscate the opcode frequency vector using another DRL based Obfuscation system and score the resulting opcode frequency vector as generated from this system with the earlier described classification system. The obfuscated variants that are detected by the detection system as benign with a probability $\mathbb{P}_{benign} > 0.5$ are used as training dataset for our de-obfuscation system.
This dataset along with the collected obfuscated opcode frequency vector from the work resulted in a set of 1612 unique opcodes. Correspondingly we have a state comprising of 1612 dimension `Action' Space with a  permissible range of [0, 10000] $ \in {\mathbb{Z}}^{1612} $.

\subsection{The design of `Action'}\label{env-action}
The reinforcement learning tasks (and hence the environments/ agents) could be broadly classified on the basis of the type of action as discrete action and continuous action RL tasks.
The high computational complexity of the continuous (and high cardinality) action space mandates use of specialized class of agents powered by special mathematical theorems \cite{DPG} that could empower both the non-deep \cite{DPG} and deep learning variants \cite{DDPG, TRPO, PPO} of reinforcement learning agents.
For each of the unique opcode in the state there are two decision/ action criteria. First is the direction of change in each, namely, to increase it or to decrease it, and second is the amount by which the increment/ decrease should occur. 
In this approach we have $\mathcal{N}_{\text{observation}} \times 2$ actions, the first $\mathcal{N}$ actions corresponding to an increase in the specific opcode frequency by a constant $\mathcal{C}_{\text{increment}}$, and the next $\mathcal{N}$ actions represent an act of decreasing the corresponding opcode frequency by a fixed amount $\mathcal{C}_{\text{decrement}}$, where, $\mathcal{C}_{\text{increment}}, \mathcal{C}_{\text{decrement}} \in \mathbb{N}$. In our implementation we have kept $\mathcal{C}_{\text{increment}} = \mathcal{C}_{\text{decrement}} = 1$. Also, since from the perspective of obfuscation, the easiest way of creating multiple obfuscation often increases the opcode frequency by adding junk code, instructions, indirect routing etc. \cite{ObfuscationTechniques}. Therefore, to mimic this effect we allow the agent's action only a \textit{net} increase in individual opcode frequency from their initial level (as in original malware). An action with a net effect of decreasing an individual opcode below its original level results in returning the same state as before the action and a commensurate reward.
We keep additional action constraints for our agent to ensure that its behavior mimics the de-obfuscation action while preserving the original functionality. Since obfuscation is mostly created by techniques which generally increases the opcode frequencies in the resulting file, therefore a good de-obfuscation system should ideally reverse this effect. So, our agent could only take actions resulting in a net reduction of any specific opcode's frequency from its original level in the obfuscated file. Also, since a negative opcode frequency is not possible, so the least it could be decreased is to zero.

\subsection{The formulation of `Reward' function}\label{env-reward}
What the agent learns and how quickly it converges is dependent upon the reward function (i.e. the reward/ penalty criteria and magnitude) to a considerable degree. Our primary objective is that the agent could alter the opcode frequency enough to substantially enhance the IDS's capability to detect it as malicious. 
Since the benign probability of the samples we selected are $\geq 0.5$ (i.e. $\mathbb{P}_{malicious\_min\_required} = 0.5$), and since the maximum possible malicious probability is 1.0 (i.e. $\mathbb{P}_{malicious\_max\_attainable} = 1.0$), we take a mid-point of these two extremes (i.e. $\mathbb{P}_{malicious}$ 0.75) as in equation \ref{eq:prob-threshold} as the preliminary target for the system. This could be stated as $(\text{given } \text{opcode} \in \mathbb{W}^{1612}, and \mathbb{P}_{(M)}=\mathbb{P}_{(Mal|IDS)})$:
\begin{multline}
    \mathbb{P}_{target}=\frac{\mathbb{P}_{(M=Certain)} + \mathbb{P}_{(M=Ambiguous)}}{2}\\
    \mathbb{P}_{target} = \frac{1}{2} (1.0 + 0.5) = 0.75 \\
    \mathbb{P} (\text{opcode}_\text{file} \mid file \subseteq \{malicious files\})>\mathbb{P}_{target}\\
    \text{or}, \mathbb{P} (\text{opcode} \mid file \subseteq \{malicious files\})>0.75
    \label{eq:prob-threshold}
\end{multline}
We penalize any resulting opcode frequency vector that has predicted probability of malicious $ \le 0.75$ and reward the ones with probability of malicious $ \geq 0.75$ proportionally. So, in each step the instantaneous reward given to the agent could be stated as equation \ref{eq:reward-function_1} (given):
\begin{multline}
    \mathbb{P} (\text{opcode} \mid file \subseteq \{malicious files\}) = \mathbb{P}_\text{malicious} \\
    \text{reward} = \mathbb{P}_\text{malicious} - 0.75
    \label{eq:reward-function_1}
\end{multline}
But this reward mechanism has a drawback that it encourages long trajectories resulting in positive rewards instead of quickly reaching a very high $\mathbb{P}_\text{benign}$. Since the `discounting-factor' $(\gamma)$ is only in the agent's control and not in environment's control, so the reward mechanism cannot take the help of lowering the discounting-factor enough so that quick high instantaneous rewards becomes more profitable than lower cumulative discounted-rewards. So, to overcome this effect, we have another (instantaneous) reward given by the environment to the agent (in addition to the one stated above), which is accrued when the agent to manage the alter the opcode frequency enough such that the file is almost unambiguously classified as malicious. This reward is high enough to easily surpass even multiple cumulative (even discounting given $\gamma < 1)$ rewards and is similar to the malicious probability for original malware variants as detected by the system. This occurs when the $\mathbb{P}_\text{benign} \geq \mathbb{P}_\text{threshold}$, where, $\mathbb{P}_\text{threshold}$ is a high/ threshold probability of malicious (say 0.99). Therefore, now the reward can be given as a step function as equation \ref{eq:reward_function} below.
\begin{equation}
    \text{reward}= 
    \begin{cases}
        \mathbb{P}_\text{malicious}-0.75,&\text{if }\mathbb{P}_\text{malicious} \\
                \text{}& \le \mathbb{P}_\text{threshold}\\
        \mathcal{R}_\text{goal},            &\text{ otherwise }
    \end{cases}
    \label{eq:reward_function}
\end{equation}
where, $\mathcal{R}_\text{goal}$ could either be a fixed constant or one dependent upon the maximum steps allowed in the episode. An episode starts with a reset of the environment. During the `reset', the environment fetches a random malware file's opcode. The episode ends when either the goal is achieved ($\mathbb{P}_{malicious} > \mathbb{P}_{threshold}$) or the maximum permissible steps for the episode is reached. Here we set $ \mathcal{R}_\text{goal} = $ Max\_Permissible\_Steps\_in\_an\_episode so that we could balance the requirements for setups with large episodes. This allows for slow but steady convergence of complex agents with too many trainable parameters. In such setups the max\_permissible\_step adaptive set in relation with the $\mathcal{R}_\text{goal}$,  such that it is always greater than any cumulative reward over even a long episode.

\section{DRL Agent(s) used}\label{rl-agent}
Given the constraints of the design of the reinforcement learning as covered in section \ref{env-action}, we have a discrete action task with a very high action space (and also state space) cardinality. Some of the most popular DRL agents for discrete action agents like the `Deep Q Networks' (DQN) \cite{DQN_Nature, DQN_Atari}, `Double DQN' (DDQN) \cite{Double_DQN}, and the `Dueling DQN' (DDQN) \cite{Dueling_DQN}. These algorithms though could manage large state-space but perform poorly for large/ continuous action space.
Deterministic Policy Gradient \cite{DPG} based deep reinforcement learning approaches like the `Deep Deterministic Policy Gradient' (DDPG) \cite{DDPG} claimed to be deliver the best in class performance on large, even continuous action-space based reinforcement learning tasks. The problem with such approaches is that their line-search based policy gradient update (as used during optimization) either proves too big for updates involving non-linear trajectory. This results in the updates overshooting the target or slower convergence. Since in the deep reinforcement paradigm non-linear gradients are quite common so algorithms based upon line-search based gradient update are not very robust and cannot provide guarantees of near monotonic policy improvements. `Trust Region Policy Optimization' (TRPO) \cite{TRPO} algorithm which is based on `trust-region' based policy updates using `Minorize-Maximization' (MM) (second order) gradient update, claims to solve this problem and provide guarantee for near monotonic general (stochastic) policy improvement even for non-linear policies like that approximated by (deep) neural networks.
Additionally, TRPO uses a mechanism called `Importance Sampling' to compute the expectancy of the policy from previous trajectories instead of only the current trajectory to stabilize the policy gradient. This method has an underlying assumption that the previous trajectory's distribution ($Q(x)$) is not very different from the current trajectory's distribution ($P(x)$). 
The policy gradient for a Stochastic Policy Gradient \cite{Stochastic_Policy_Gradient} method and associated algorithms like Actor Critic \cite{A3C} is given as equation \ref{eq:SPG}:
\begin{equation}
  \nabla_\theta (J_\theta) = \mathbb{E}_{\tau \sim \pi_\theta (\tau)} [\nabla_\theta \log \pi_\theta(\tau) r(\tau)]  
  \label{eq:SPG}
\end{equation}
In equation \ref{eq:SPG}, the trajectory $\tau$ over which the samples for computing expectancy is gathered (to update the gradient $\nabla$ of the policy-value-function J) is the same (current) trajectory of the policy as used in the policy $\pi$ (parameterized over $\theta$).
But in the case of TRPO using importance sampling and the past trajectory for sampling, this policy-value-function update looks as equation \ref{eq:TRPO-PG} below:
\begin{equation}
    \begin{aligned}
      \nabla_{\theta\prime} (J_\theta\prime) =  
      \mathbb{E}_{ \tau \sim \pi_\theta (\tau)} [\sum_{t=1}^T\nabla_\theta\prime \log \pi_\theta\prime (\prod_{t\prime=1}^t \frac{\pi_{\theta\prime}}{\pi_{\theta}}) (\sum_{t\prime=t}^T r)]  
    \end{aligned}
    \label{eq:TRPO-PG}
\end{equation}
To avoid too large changes in gradient, a penalty needs to be added to equation \ref{eq:TRPO-PG} to make the optimization more monotonic. With this penalty combined with the use of the advantage, the optimization function is given as equation \ref{eq:TRPO-Optimization-Beta}:
\begin{equation}
    \begin{aligned}
      \max_{\theta} \hat{\mathbb{E}_{t}} [\frac{\pi_{\theta}(a_t | s_t)}{\pi_{\theta old}(a_t | s_t)}\hat{A_t} - \beta KL[\pi_{\theta old}(.|s_t),\pi_{\theta}(.|s_t)]]  
    \end{aligned}
    \label{eq:TRPO-Optimization-Beta}
\end{equation}
The problem with the $\beta$ based penalty as in equation \ref{eq:TRPO-Optimization-Beta} is that it is difficult to choose a single value of $\beta$ that aligns well to different types of problems, or even for a single problem as the learning progress. Therefore to resolve this issue, TRPO instead of using $\beta$ based penalty, uses KL based constraints as shown in equation \ref{eq:TRPO-Optimization-Constraint}, thus requiring a second order optimization solution.
\begin{equation}
    \begin{aligned}
      \max_{\theta} \hat{\mathbb{E}_t} [\frac{\pi_{\theta}(a_t | s_t)}{\pi_{\theta old}(a_t | s_t)}\hat{A_t}] \\
      \text{subject to }\hat{\mathbb{E}_t}[KL[\pi_{\theta old}(.|s_t),\pi_{\theta}(.|s_t)]] \leq \delta.
    \end{aligned}
    \label{eq:TRPO-Optimization-Constraint}
\end{equation}
In TRPO, the second order gradient update computation is complicated and also very expensive, and hence for real size-able tasks it is seldom use. The `Proximal Policy Optimization' (PPO) \cite{PPO} algorithm instead of using a constrained form of solution (as shown in equation \ref{eq:TRPO-Optimization-Constraint}), clips the surrogate objective to ensure that the updates are not unconstrained. This is as given in equation \ref{eq:TRPO-Optimization-Surrogate}.
\begin{equation}
    \begin{aligned}
      L^{CPI}(\theta) = \hat{\mathbb{E}_t} [\frac{\pi_{\theta}(a_t | s_t)}{\pi_{\theta old}(a_t | s_t)}\hat{A_t}]
                      = \hat{\mathbb{E}_t} [r_t(\theta)\hat{A_t}].
    \end{aligned}
    \label{eq:TRPO-Optimization-Surrogate}
\end{equation}
The original surrogate objective $L^{CPI}$ for TRPO as discussed in equation \ref{eq:TRPO-Optimization-Surrogate}. This in the `clipped' form could be reformulated as equation \ref{eq:PPO-Optimization-Clipped}, where, $\epsilon$ is a hyper-parameter. The default value of $\epsilon$ is set to 0.2. 
\begin{equation}
    \begin{aligned}
      L^{CLIP}(\theta) = \\
      \hat{\mathbb{E}_t} [\min(r_t(\theta)\hat{A_t},clip(r_t(\theta),1-\epsilon,1+\epsilon)\hat{A_t})].
    \end{aligned}
    \label{eq:PPO-Optimization-Clipped}
\end{equation}
Another suggested variant of the PPO algorithm is based upon adaptive $\beta$ penalty as given in equation  \ref{eq:PPO-Adaptive}. But in various experiments, the clipped penalty form of equation \ref{eq:PPO-Optimization-Clipped} performed better than the adaptive penalty form of equation \ref{eq:PPO-Adaptive}, and hence we use the clipped form in our system.
\begin{equation}
    \begin{aligned}
        \beta =
        \begin{cases}
             \beta/2, & \text{if } d \le d_{targ}/1.5\\
             \beta\times2, & \text{if } d \ge d_{targ}\times1.5\\
        \end{cases}
    \\ where, d = \hat{\mathbb{E}_t} [KL[\pi_{\theta}(. | s_t), \pi_{\theta old}(. | s_t)]]
    \end{aligned}
    \label{eq:PPO-Adaptive}
\end{equation}
The PPO algorithm works similar to TRPO and is much easier to compute as it uses a linear variant of the gradient update called the `Fisher Information matrix' (FIM). In equation \ref{eq:TRPO-PG}, the trajectory is sampled from the policy as it existed in previous time (t) as $\pi_{\theta} = Q(x)$. The expectancy over such collected samples are used to update the policy at next time step $(\pi_{\theta\prime} = P(x))$. When the ratio of expectancy over the two trajectory distributions ($\frac{P(x)}{Q(x)}$) vary significantly as in the case of linear gradient update in PPO, the previously stated assumption may not hold, leading to high variance in policy updates. To avoid this there are two methods that the PPO algorithm recommends. The first one use a `Adaptive KL Penalty' and the second one use `Objective Clipping'.
As per the original PPO paper \cite{PPO}, the `Objective Clipping' variant, with the clipping factor $\epsilon = 0.2$ provided the best result. We use similar mechanism in our experiment. In this mechanism if the probability ratio between the two trajectory's policies is not in the range $[(1-\epsilon),(1+\epsilon)]$ the `estimated advantage' is clipped.
We use the Proximal Policy Optimization algorithm (PPO) \cite{PPO}. PPO is an improvement over Trust Region Policy Optimization algorithm (TRPO) \cite{TRPO}. 
The deep learning model that we use for the PPO algorithms actor and critic network comprise of 2 hidden layers each, with each hidden layer having 64 neurons and a `tanh' activation function.

\section{Experiments and Results}\label{results}
\begin{figure*}[htb]
    \centering
    \includegraphics[width=\linewidth, height=1.8in]{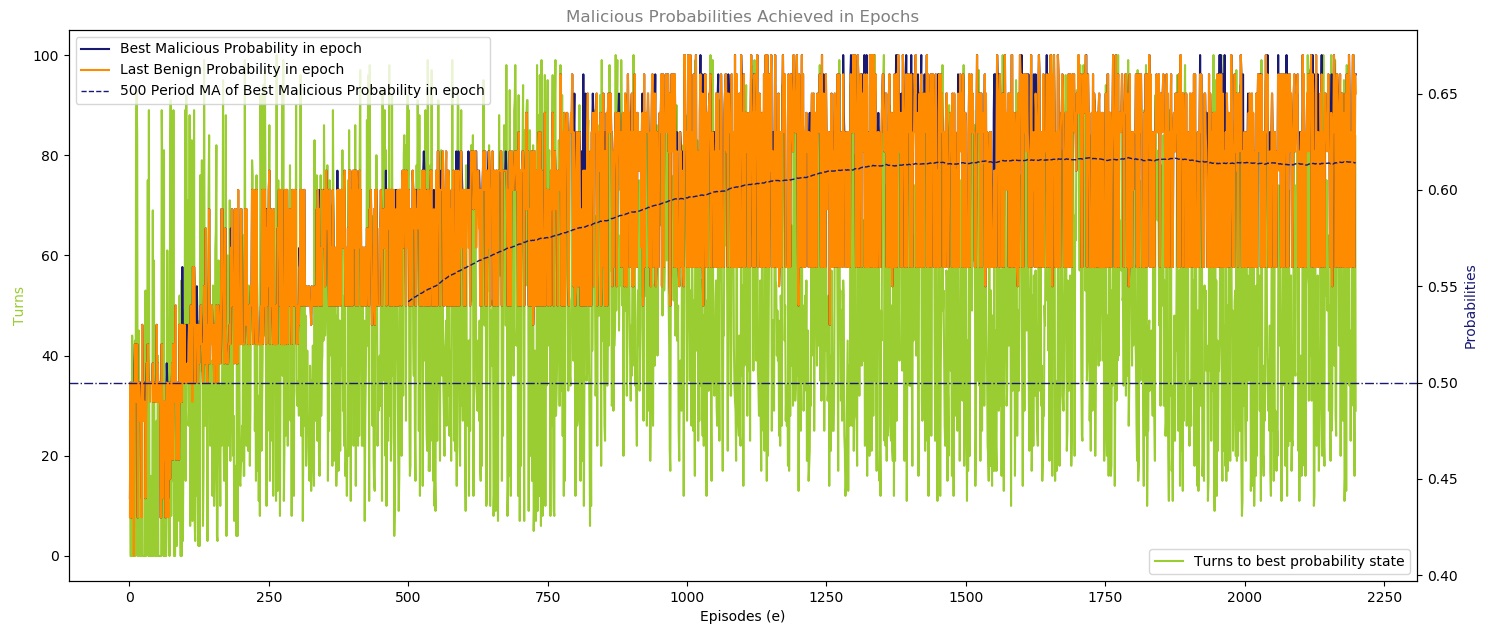}
    \caption{Training Statistics - Malicious Probability detected by the IDS after De-Obfuscation across episodes.}
    \label{fig:training_statistics_agent_1_malicios_prob_episodic}
\end{figure*}

\begin{figure*}[htb]
    \centering
    \includegraphics[width=\linewidth, height=2in]{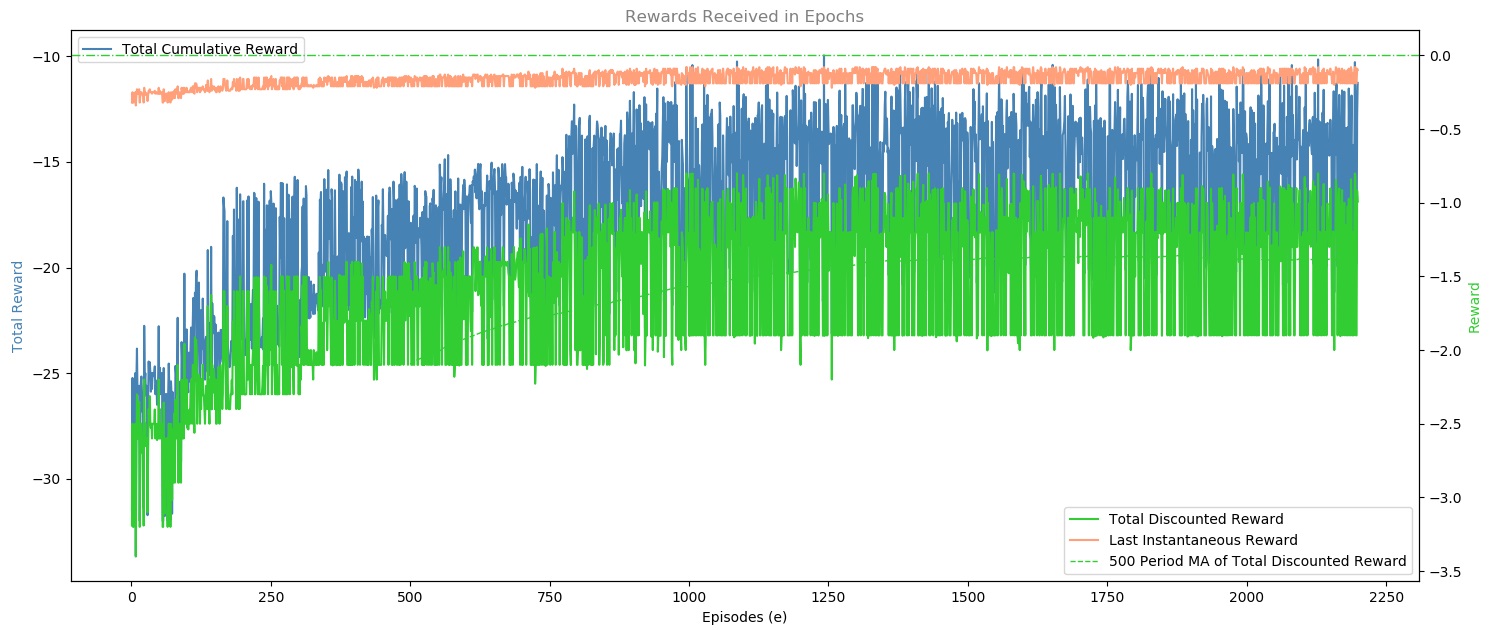}
    \caption{Training Statistics - Total Instantaneous and Discounted rewards accrued by agent across episodes.}
    \label{fig:training_statistics_agent_1_total-rewards}
\end{figure*}

\begin{figure}[htb]
    \centering
    \includegraphics[width=\linewidth]{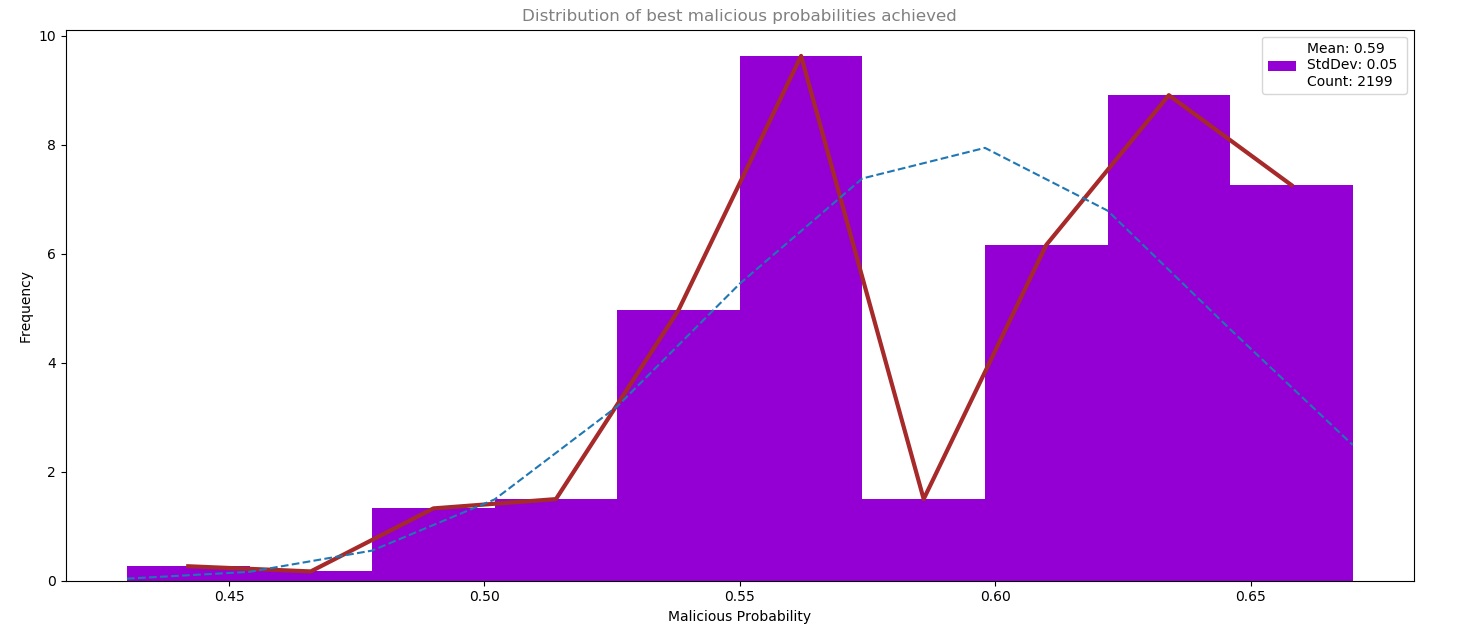}
    \caption{Training Statistics - $\mathbb{P}_{malicious}$ as detected by the IDS for the de-obfuscated version of the malware by the agent.}
    \label{fig:training_statistics_agent_1_malicious-prob-hist}
\end{figure}

We conducted over 2000 (episodic) experiments where the PPO algorithm based Deep Reinforcement Learning (DRL) agent would attempt to de-obfuscate a malicious file and validate if the associated IDS could then detect the de-obfuscated version of the obfuscated malicious file.
Each such experiment is constituted of an episode consisting of several steps. In the first step the environment extracts a random new obfuscation of any malicious file from the opcode frequency vector feature repository (as generated by the ADRLMMG system) and sends it to the agent to process it. The agent then alters the frequency of one of the opcode in each subsequent step and the environment correspondingly rewards the agent as per the mechanism described in section \ref{env-reward}.
For each episode, we record the malicious probability that was finally reported by the existing IDS on the de-obfuscated version of the opcode frequency vector as generated by the agent in the final step of every episode (each episode starts with a new obfuscated feature-vector). We use the existing IDS as-is without altering or retraining it. These probability $\mathbb{P}_{malicious}$ trends across episodes are reported in the plot in figure \ref{fig:training_statistics_agent_1_malicios_prob_episodic}. As shown in this figure, the detected $\mathbb{P}_{malicious}$ (moving average) has crossed the critical point of $\mathbb{P}_{malicious} > 0.5$ very early and has reached $\mathbb{P}_{malicious} > 0.6$ in around 2000 episodes.
In each step of every episode the agent gets some instantaneous rewards. The rewards received earlier in an episode could be discounted by a discounting factor to give more importance to more recent awards. The total of all instantaneous rewards or the discounted rewards indicates how well the agent is learning to achieve its objectives as converted into the rewards using the defined reward function for the agent. As shown in figure \ref{fig:training_statistics_agent_1_total-rewards}, as the episodes progress the agent is able to accrue higher total instantaneous rewards and discounted rewards (left y-axis) indicating that the agent is able to effectively learn the desired policy. Also, the last instantaneous reward for most of the episodes (right y-axis) is high indicating that the episodes are ending in successful detection of the transformed opcode frequency features as malicious by the IDS.
The total rewards may just be a single indicator of how well the agent is learning the mathematical abstraction of the tasks formulated into the reward function. Additionally, to ensure that the agent's learning is aligned well with the desired outcome that the IDS should be able to detect the malicious file correctly after de-obfuscation, we present the histogram of all the final $\mathbb{P}_{malicious}$ detection probability of the de-obfuscated feature (opcode frequency) vector of the obfuscated/ metamorphic malware file by the existing IDS without re-training or modifying the IDS in figure \ref{fig:training_statistics_agent_1_malicious-prob-hist}. As shown in this figure the mean malicious probability ($\mathbb{P}_{malicious\_mean}$) was uplifted to $\sim 0.6$ (where, $\mathbb{P}_{malicious\_initial} \in [0.0, 0.5)$), indicating that the IDS could now effectively detect the generated de-obfuscated variants as malicious with high $\mathbb{P}_{malicious}$ probability.

\section{Discussion} \label{discussion}
\begin{figure}[htb]
	\centering
    \includegraphics[width=\linewidth]{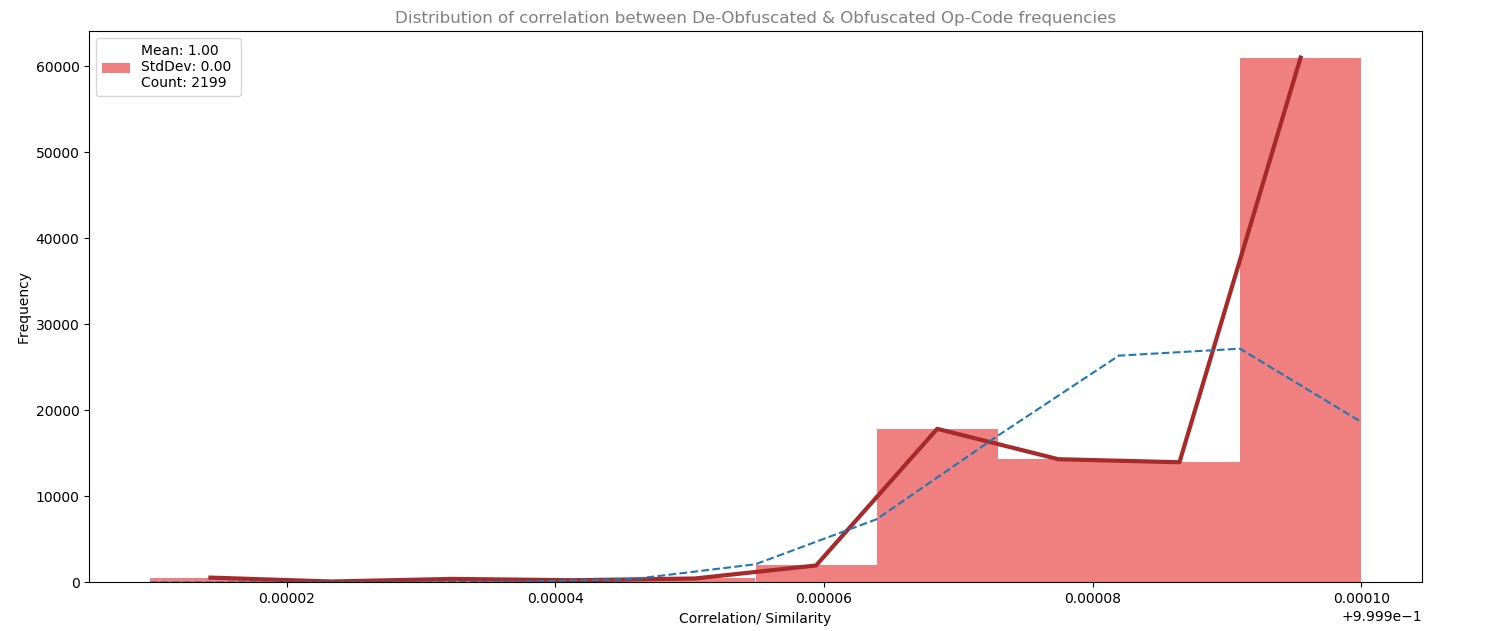}
    \caption{Training Statistics - Histogram of Similarity of the feature-vector of the De-Obfuscations produced by the agent with the Obfuscated Variant feature-vector of the Malware as given as input to the agent.}
    \label{fig:training_statistics_agent_1_histogram-similarity-obfuscated}
\end{figure}

\begin{figure}[htb]
    \centering
    \includegraphics[width=\linewidth]{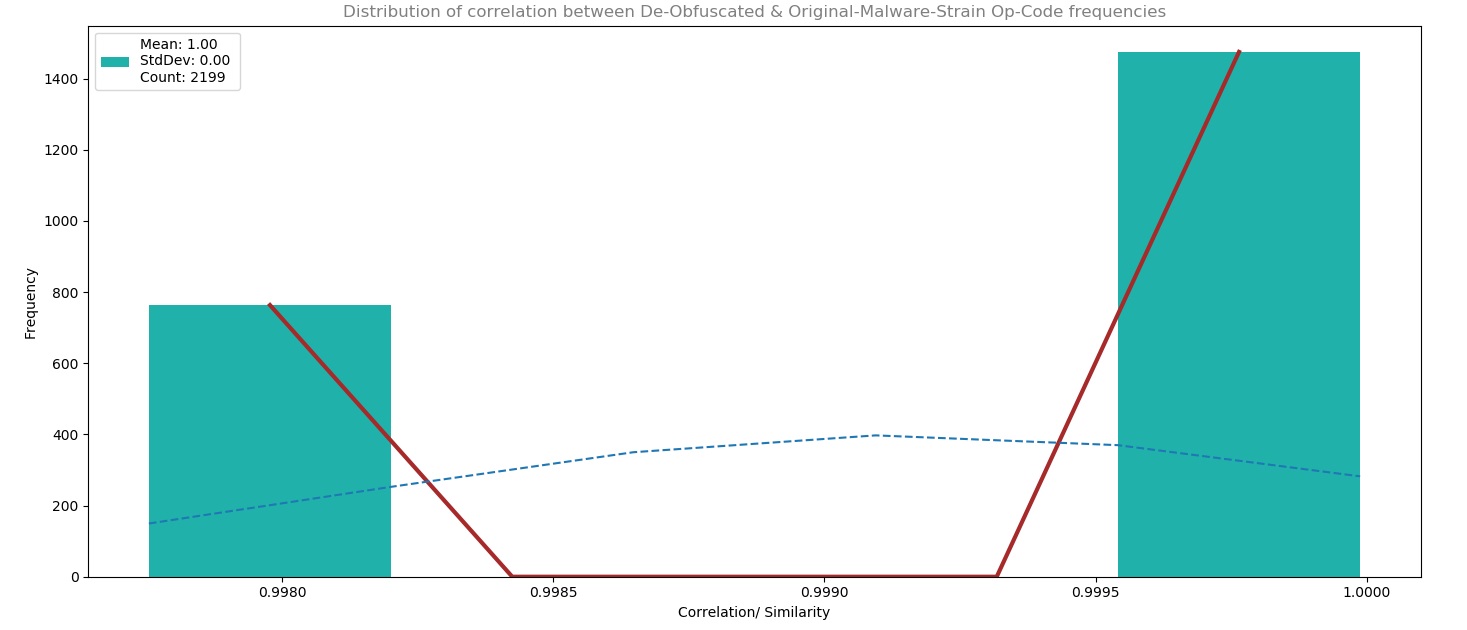}
    \caption{Training Statistics - Histogram of Similarity of the feature-vector of the de-obfuscations produced by the agent with the feature-vector of the Original Malware (that was not exposed to the agent).}
    \label{fig:training_statistics_agent_1_histogram-similarity-original-malware}
\end{figure}

As covered in section \ref{results}, the IDS is able to detect the obfuscated malicious file after the resultant de-obfuscation by the DRLDO agent as malicious, which helps probably in evading a `zero-day attack' by a metamorphic/ obfuscated variant of an existing malware. But still we are not sure whether the agent is actually learning to de-obfuscate/normalize the obfuscated variant of the malware or learning some non-reproducible tricks or exploiting some trivial loopholes to help the IDS raise the $\mathbb{P}_{malicious}$ of some specific variant. To ascertain that the agent is actually learning to de-obfuscate the metamorphism in the malware we compute and record the correlation similarity of the final output (de-obfuscated) feature-vector (opcode frequency vector) as produced by the agent with the opcode feature-vector of the obfuscated variant that the agent received as an input. We also compute the correlation similarity of the final output feature-vector with that of the original malware's corresponding feature-vector. 
Figure \ref{fig:training_statistics_agent_1_histogram-similarity-original-malware}, and \ref{fig:training_statistics_agent_1_histogram-similarity-obfuscated} shows the histogram of the overall correlation similarities between the feature-vector of the de-obfuscations produced by the agent with the feature-vector of the original malware's and between feature-vector of the de-obfuscations produced by the agent with that of its obfuscated variant's feature-vector as submitted to the agent respectively.
We use Pearson product-moment correlation coefficients between the opcode vectors to generate these similarities. The correlation is taken from the correlation matrix R, whose relationship with the co-variance matrix, C, is as given as $R_{ij} = \frac{ C_{ij} }{ \sqrt{C_{ii}*C_{jj}} }$.
Another interesting observation is related to the resultant output's opcode frequency vector's correlation similarity. We measure two types of correlations, the first is the similarity between the de-obfuscated opcode frequency vector as created by the DRLDO system and that of the obfuscated opcode frequency vector as provided as an input to the DRLDO system, and the second similarity is measured between the de-obfuscated opcode frequency vector as generated by the DRLDO system and that of the opcode frequency vector of the original malware. The opcode frequency vector of the original malware is not known to the DRLDO system and we extract it from the metadata created by the ADRLMMG system which produced these obfuscations. The feature-vector of the de-obfuscated variants as produced by the DRLDO system were quite similar to ($\geq 0.99$) their original malware's feature-vector and very dissimilar to ($\leq 0.001$) their obfuscated variants as submitted to the DRLDO system. This observation establishes that the desired enhancement in the detection probability of $\mathbb{P}_{malicious}$ of these feature-vectors after processing them via the DRLDO system is actually due to the probable de-obfuscation carried out by the DRLDO system in which it removed some of the additional opcode/ instructions that were inserted in the original malware to evade its detection by the IDS and the results are unlikely to be because of any trivial and non-reproducible trick that the agent might have unintentionally learnt.
The insights from these observations are very significant especially since we never exposed the original malware strain's opcode frequency vector to the DRLDO system. Attaining such high similarity with the original malware's opcode frequency indicates that the resultant de-obfuscations thus created by our system could not only be now detected by most existing IDS as malicious, but it could also be identified that the incoming malware is an obfuscated variant of one of the existing malware variants that the IDS has in its training repository. So, besides enhancing binary IDS that could just detect whether a file is benign or malicious the de-obfuscations created from our system is also compatible with and would produce correct results with a multinomial IDS that also detects the family of the malware variants. Alternatively in a binary IDS subsequently by using the similarity between the outputs of the DRLDO system with the stored feature-vectors of the existing malware variants in the system's repository the family of the obfuscated malware could be identified thus enhancing the insights generated from the detection system.
The above observations on the similarity with the original malware strain also indicates that following de-obfuscation, the file size and opcode frequency distribution does not change substantially. This has other significant implications as this would also prevent any malware prediction probability creep/ enhancement even when subjected to any IDS which first segregates the files into different categories on the basis of either their file size \cite{ashu_filesize} or on the outcomes of machine learning methods like clustering  \cite{hemant_clustering} before scoring/ predicting them for their maliciousness, to enhance their respective prediction/ detection effectiveness/ accuracy. 

\section{Conclusion}\label{conclusion}
We designed and developed an advance Deep Reinforcement Learning based system named DRLDO that could learn how to de-obfuscate and normalize a metamorphic (or otherwise obfuscated) malware. Unlike some other systems that could work only at the binary level and hence the transformations from these are intractable and non-functionality-preserving, the DRLDO is the first system that could perform de-obfuscations at the opcode-sequence level. Additionally, the DRLDO system offers unique advantages as it does not mandate any change in the IDS' classification-system and does not even require a re-training of the classifier. Thus, the DRLDO system could be easily retrofitted into an existing IDS setup.
The experiments conducted with the DRLDO system, and the corresponding results obtained, proves that PPO algorithm based DRL agents, as used in the DRLDO system, could be effectively trained using our custom-developed RL environment. 
The so trained DRL agents could effectively de-obfuscate the (opcode-sequence) feature-vector of an incoming obfuscated malware. The resulting transformed feature-vector could be correctly detected by an existing IDS with a detection probability of up to $0.6$ for previously un-detectable obfuscated intrusions. In the entire process no re-training, re-configuration or re-calibration of the IDS is required. 
\par
Thus, the DRLDO system could effectively provide an existing IDS the augmented capabilities of defense against (even multiple-simultaneous) attack from metamorphic variants of existing malware. Doing so, the DRLDO system can enhance an IDS with unique defensive capabilities against any probable `zero-day attack' by a metamorphic attack from obfuscated variants of an existing malware.

\bibliography{references}
\bibliographystyle{dsj}
\end{document}